\documentclass[%
 aip,
 reprint,%
]{revtex4-1}
\usepackage[version=4]{mhchem}
\usepackage{graphicx}

\usepackage[english]{babel}
\usepackage{bm}

\usepackage{lineno}

\usepackage[dvipsnames]{xcolor}

\newcommand{\response}[1]{#1}

\begin{document}

\title{Anatomy of Leadership in Collective Behaviour}

\author{Joshua Garland}
\email{Joshua@santafe.edu}
\affiliation{Santa Fe Institute, Santa Fe, NM 87501}

\author{Andrew M.\ Berdahl}
\affiliation{Santa Fe Institute, Santa Fe, NM 87501}
\affiliation{School of Aquatic \& Fishery Sciences, University of Washington, Seattle, WA 98195}

\author{Jie Sun}
\affiliation{Department of Mathematics, Clarkson University, Potsdam, NY 13699}
\affiliation{Department of Physics, Clarkson University, Potsdam, NY 13699}
\affiliation{Department of Computer Science, Clarkson University, Potsdam, NY 13699}
\affiliation{Clarkson Center for Complex Systems Science, Clarkson University, Potsdam, NY 13699}

\author{Erik M.\ Bollt}
\affiliation{Department of Mathematics, Clarkson University, Potsdam, NY 13699}
\affiliation{Clarkson Center for Complex Systems Science, Clarkson University, Potsdam, NY 13699}
\affiliation{Department of Electrical and Computer Engineering, Clarkson University, Potsdam, NY 13699}

\date{\today}

\begin{abstract}
Understanding the mechanics behind the coordinated movement of mobile animal groups \response{(collective motion)} provides key insights into their biology and ecology, while also yielding algorithms for bio-inspired technologies and autonomous systems.
It is becoming increasingly clear that many mobile animal groups are composed of heterogeneous individuals with differential levels and types of influence over group behaviors. 
The ability to infer this differential influence, or leadership, is critical to understanding group functioning in these collective animal systems. Due to the broad interpretation of leadership, many different measures and mathematical tools are used to describe and infer ``leadership", e.g., position, causality, influence, information flow. 
But a key question remains: which, if any, of these concepts actually describes leadership? 
We argue that instead of asserting a single definition or notion of leadership, the complex interaction rules and dynamics typical of a group implies that leadership itself is not merely a \response{binary classification (leader or follower)}, but rather, a complex combination of many different components. In this paper we develop an anatomy of leadership, identify several principle components and provide a general mathematical framework for discussing leadership. \response{With the intricacies of this taxonomy in mind we present a set of leadership-oriented toy models that should be used as a proving ground for leadership inference methods going forward.} 
We believe this multifaceted approach to leadership will enable a broader understanding 
of leadership and its inference from data in mobile animal groups and beyond.
\end{abstract}


\maketitle

\begin{quotation}
When observing the collective motion of animal groups (e.g., schooling, herding, or flocking), 
an immediate question is, what is the leadership structure? 
Who (if anyone) is in charge and who is following, and does such structure stay the same or change over time? 
Recent technological advances in image processing and animal-mounted sensors make it possible to record the simultaneous movement trajectories of every animal in a group.
Such abundance of data makes the present a promising  time to progress in understanding leadership structure in mobile animal groups. 
Despite the availability of data and the central importance of understanding leadership in collective motion, there is surprisingly little explicit mathematical description or even a consistent and well-defined approach to this subject.
Here, as a first step toward addressing this deficiency, we construct a framework for inferring leadership in collective motion.
We review various sources and characteristics of leadership to provide an anatomy  and a language for describing the multifaceted aspects of leadership across a variety of animal societies.
We then present a suite of leadership-focused toy models, which can be used as a proving ground for any proposed leadership inference method, before being naively applied to (empirical) data.
Together, this lays the groundwork for a principled exploration of a perennial question:
how is control of a collective system distributed? Such understanding will not only contribute to the ecology and conservation of group-traveling species, but will also aid in the design of control algorithms for emerging distributed technologies.
\end{quotation}

\section{Overview}
\label{sec:intro}

Mobile animals groups (e.g., flocks, herds, schools, swarms) are ubiquitous in nature. 
In such collective systems, the interactions between individuals may be as important as characteristics of the individuals themselves\cite{vicsek2012collective}. 
Insight into these interactions and their impact on the group dynamics is of fundamental importance for our understanding of both the ecology of these systems\cite{westley2018collective} as well as design and control principles underlying general complex systems \cite{anderson1972more}.

A key challenge in the study of collective animal behavior is  understanding how groups of organisms make decisions as a whole\cite{conradt2005consensus}, for example about where\cite{berdahl2018colnavrev} or when\cite{helm2006sociable,berdahl2017social} to go.
Group decision-making processes range from despotic to shared\cite{conradt2003group},
although even in systems with shared or distributed decision making there are likely inter-individual differences (e.g., sex, rank, personality, size, nutritional state, informational state) that produce asymmetry in influence.
Models suggest that such heterogeneity is potentially important to group-level dynamics\cite{couzin2005effective,delgado2018importance}, but inferring differential influence and leadership from empirical data, though often attempted, is an open challenge.
As we elaborate in some detail in this paper, a key step toward tackling this challenge lies in the recognition that the notion of leadership is not merely a simple, unidimensional concept. Instead, a rich palette of different types and forms of leadership often coexists, even for the very same system. Thus, we argue that a precursor step to the ``correct'' inference of leadership is the clarification of what (type of) leadership is sought of. Without such, any inferred leadership can potentially be deemed inappropriate.

The need to distinguish between the {\it definition} and {\it inference} of leadership is standing out as a central problem partly because of the acceleration of technical progress that enabled collection of ``big" data. For example, 
new technologies to collect the simultaneous trajectories of all members of a mobile animal group\cite{hughey2018challenges}, along with increases in computing power, make the near future a fruitful time to meet this challenge. 
Will having large amount of real-world data alone be sufficient to address questions about leadership, or do we (still) need conceptual advances?
As recently reviewed by Strandburg-Peshkin et al.\cite{strandburg-Peshkin2018inferring}, most efforts to infer leadership have used position within a group\cite{smith2016leadership,lewis2011highly,brent2015ecological,jacoby2016inferring} (e.g., leaders are assumed to be at the front), initiator-follower dynamics\cite{strandburg2015shared,leca2003distributed} or time-delayed directional correlations \cite{nagy2010hierarchical,akos2014leadership,jiang2017identifying,watts2017validating,ye2015distinguishing}. Information theoretic measures provide additional, potentially more powerful and less subjective, tools to infer leadership and influence\cite{lord2016inference,butail2016model,sun2015causal}. However, a central viewpoint of this paper is that any measurement of leadership needs to start by clarifying the particular type or form of leadership one is after. 
Without such clarification, the ``leadership" resulted from the application of any inference method can be subject to misinterpretation, and perhaps more seriously, lead to fundamentally flawed conclusions about the interaction mechanisms of an animal system.

To illustrate the many facets of leadership and thus the need to distinguish between its definition and measurement,
consider, for example, the case of migrating caribou.  Older, more experienced individuals are thought to guide the migration-scale  movements\cite{eyegetok2001thunder}, however, pregnant or nursing females might have increased nutritional requirements\cite{barboza2008allocating} and thus guide movements along that path towards habitat with better forage opportunities\cite{conradt2009leading}. 
Therefore, who is leading depends on the time- and length-scale of the movements considered. Additionally, for some populations fall migration coincides with the rut, so mating behaviors drive social interactions: a dominant male may attempt to herd females or drive other males away. Such a male is certainly influential, but perhaps should not always be considered a leader, at least in the context of the migration. Finally, whether or not an individual is a leader might depend on who (or which group) one is considering as a potential follower. A nursing (and thus infertile) female might be ignored by the 
libidinous male, but will be closely followed by her calf\cite{torney2020inferring}.
Because there are many scales and types of influence/leadership, we argue that one should begin such explorations with a clear question and select analytical methods to match.

The central goal of this paper is to develop a formal language and multifaceted framework for defining and (potentially) inferring the many aspects of leadership. In addition, we aim to provide a set of leadership-oriented toy models to serve as a proving ground for leadership inference methods. Thus our work here offers a practical language and set of tools for researchers hoping to match questions about leadership with the appropriate methods while avoiding potential pitfalls. We hope that the combination of mathematical rigor, biological intuition, together with several real and synthetic examples will make our framework accessible and interesting to both biologists and applied mathematicians.

\section{General Mathematical Framework}
\label{sec:math-frame}

To capture various forms of leadership, consider dynamics of individuals (with potential interactions among them via a network) together with dynamics of the group determined by the individuals, modeled by the general form of ODEs:
\begin{equation}\label{eq:general} 
\begin{cases}
\dot{\bm{x}}_i = \bm{f}(S_{i1}(t)\bm{x}_1(t),\dots,S_{in}(t)\bm{x}_n(t);\bm{\mu}_i(t);\bm{\xi}_i(t)),\\
\bm{y}(t) = \bm{h}(\bm{x}_1(t),\dots,\bm{x}_n(t)).
\end{cases}
\end{equation}
In this general model class, $\bm{x}_i(t)$ represents the state of the $i$-th individual at time $t$ ($i=1,\dots,n$), $S=[S_{ij}(t)]_{n\times n}$ is the (time-dependent) adjacency matrix (also known as the \textit{sociality} matrix) of a network encoding the structure of interactions, where $S_{ij}\neq0$ if it is possible for $j$ to (directly) impact the state of $i$. Furthermore, $\bm{\mu}_i(t)$ denotes the parameter (vector) associated with $i$, and $\bm{\xi}_i(t)$ is noise. \response{Here a ``parameter'' can be anything that describes the heterogeneity of the individuals in the group. For example, in the Viscek model\cite{vicsek1995novel}, the parameter $\bm{\mu}_i$ can represent the preferred direction an individual takes, or it can also be used to represent the speed of an individual that might differ from one to another, or both by associating a parameter vector to each individual.} The function $\bm{f}$ models how the dynamics of each individual depends on their own state and parameter(s), the state of others in the network, and noise. Finally, the state of the group, \response{$\bm{y}(t)$}, is determined by the state of the individuals through the function $\bm{h}$; for example, taking $\bm{h}(\bm{x}_1(t),\dots,\bm{x}_n(t))=\frac{1}{n}\sum_{i=1}^{n}|\bm{x}_i(t)|$ defines the group state as the average of the individuals states.

A separate and complementary perspective is to model/represent the individual and group dynamics as a multivariate stochastic process, focusing on stationary variables $\bm{X}_i(t)$ and $\bm{Y}(t)$. From this perspective, the relationship between the group variable and  the variables are encoded in the conditional distribution function
\begin{equation}\label{eq:general2}
p(\bm{y}(t)|\bm{x}_1(t^-),\bm{x}_2(t^-),\dots,\bm{x}_n(t^-)),
\end{equation}
where $t^-=(t-\tau,t)$ denotes time history of the system, taking into account a time lag of $\tau\in(0,\infty)$.  

We point out that there is intimate connection between a dynamical system [such as one defined by Eq.~\eqref{eq:general}] and a stochastic process, generally through an underlying (ergodic) measure~\cite{bollt2013applied}, where the uncertainty associated with the state of the variables is generally related to the distribution of initial conditions and noise in addition to the coupled dynamics. %
For a deterministic system, the randomness initiates exclusively from (experimental) imperfection of choosing and determining the initial condition, and the evolution of uncertainty can be treated as a stochastic process. Thus entropy methods are naturally associated even with otherwise deterministic dynamical systems Eq.~(\ref{eq:general}) in terms of the associated stochastic process.

From the stochastic representation~\eqref{eq:general2} of the dynamics, we can define an individual's (observed) influence on the group using various forms of conditional mutual information. For example, the (unconditioned) mutual information (MI)
\begin{equation}
I(\bm{x}_i(t^-);\bm{y}(t))
\end{equation}
measures the apparent influence of $i$ on the group, aggregated over both direct and indirect factors. On the other hand, after factoring out indirect factors, the ``net'' influence of $i$ on the group can be measured by the conditional mutual information (CMI) 
\begin{equation} \label{eq:influence}
I(\bm{x}_i(t^-);\bm{y}(t)|\bm{x}_{\bar{i}}(t^-)),
\end{equation}
where $\bar{i}=\{1,\dots,i-1,i+1,\dots,n\}$.
\response{As suggested recently by James et al.\cite{james2016information}, Eq.~\ref{eq:influence} may not capture influence entirely, therefore care should be taken when quantifying net influence in this way.}

Note that Eq.~\eqref{eq:general} itself does not uniquely determine the distribution in Eq.~\eqref{eq:general2}, due to the possibly different states/trajectories the system can follow depending on initial conditions, parameters, and other factors; unique ergodicity and fixed parameters are possible assumptions if we wish to discuss uniqueness. Equation~\eqref{eq:general} can be interpreted as modeling the possible interactions among the individuals, although these interactions may or not not be realized in a particular setting depending on the states the system operates in; on the other hand, the PDF in Eq.~\eqref{eq:general2} encodes (intrinsic) dependence between the group variable and those of the individual variables without necessarily matching the structural information in Eq.~\eqref{eq:general}, even if such dependence comes from dynamics of Eq.~\eqref{eq:general}.

Next, we distinguish between intrinsic states of the system versus observed states, as a key aspect in mathematical interpretation of any process, including group roles of leadership, is the concept of measurement of observables, from the underlying process.  
In fact, the concept of leadership and information flow can be dramatically obscured depending on the details of the observables (extrinsic variables) relative to the underlying system (intrinsic variables).
We use $\hat{\bm{x}}_i(t)$ to represented the observed state regarding $\bm{x}_i(t)$, and similarly, $\hat{\bm{y}}(t)$ for the observed state regarding $\bm{y}(t)$. We represent the observations over a finite time window, producing observational data 
\begin{equation}
	\{\hat{\bm{x}}_i(t);\hat{\bm{y}}(t)\}_{t\in\mathcal{I}}.
\end{equation}

Proper characterization and interpretation of leadership requires the (subjective) identification of a ``reference frame", namely, 
choosing the (observable) variables, groups, as well as time and spacial scales. That is, we argue that defining such a frame needs to include making at least the following three choices:
\begin{enumerate}
\item Variables (e.g., position, velocity, acceleration, direction of motion or some combination of these). Depending on the choice of variables, different types of leadership can be defined and (potentially) identified.
\item \response{Temporal resolution and time lag. What is the temporal resolution of the actions of interest (e.g., seconds, days, or years)? Additionally, there is an issue of time lag. How far into the future is an action thought to have potential impact? 
If the time lag is larger than the time-scale of the typical response to an individual's action, then each individual will appear to have a similar random influence on the others. On the other hand, too small of a time lag might prevent detection of the (time-delayed) dynamics of the group in response to an individual's actions.
}
\item Definition of a group and what it represents. For example, a group can contain everyone within a spatial domain, or can be a certain class of individuals based on age, gender, etc.
\end{enumerate}

\medskip
\section{Principle Components of Leadership}
\label{sec:pcl}

\response{In broad terms, we define leadership as an individual having asymmetric potential to impact the trajectory of agents in the group.} As we explore below, the source of this asymmetrical impact or influence may be due to group structure, individual information or emerge from social interaction rules alone. Further, the distribution and time and length scales of the resulting leadership may vary considerably. In this section we construct a series of informative classifications which we will refer to as the \textit{components} of leadership. We further divide these components into \textit{sources} and \textit{characteristics} of leadership.

\subsection{Sources of leadership}

\noindent\textbf{Structural Leadership.}
Structural leadership encompasses a wide range of leadership which fundamentally relies on the structure of the animal society. \response{This structure could be  an explicit dominance hierarchy, or more subtly due to unequal social influence due to semi-persistent traits (e.g., age, gender, reproductive status). Depending on the particular taxa, the driving mechanism for such asymmetric interactions differ and deriving such a mechanism is not the purpose of this manuscript. For simplicity, we assume all of this rich societal structure has been {\it pre-encoded} in the sociality matrix defined in Eq.~\ref{eq:general}. In particular, $S_{ij}\neq 0$ if and only if $j$ has the capacity to lead $i$ directly. Where ``capacity to lead'' is defined by the particular society.} 

\response{To formalize this component of leadership, let $\mathcal{G}$ be the directed graph associated with the sociality matrix $S$, where there exists an edge from $j$ to $i$ if $S_{ij}\not=0$. For each node $\ell\in \mathcal{G}$, denote the reachability set of node $\ell$ as $\mathcal{F_\ell}$. In particular, node $k$ is a member of $\mathcal{F_{\ell}}$ if there exists a directed path from $\ell$ to $k$ in $\mathcal{G}$. If $\mathcal{F_\ell}\not= \emptyset$ then $\ell$ is defined to have capacity for \textit{structural leadership}. We define the set of individuals with non-zero capacity to exhibit structural leadership (have a nonempty reachability set on the sociality matrix) as $\mathcal{L}$.
Of course, the degree to which an individual is a structural leader exists on a continuum. 
Quantifying the strength of such leadership is a highly non-trivial and potentially system-specific task (e.g., \cite{flack2006encoding,brush2013family,de2017physical}). However, to first order, individuals with many individuals downstream of them and fewer individuals upstream of them in the sociality matrix will tend play a stronger leadership role, or at least have the potential to do so.}

\response{In our caribou example from the Introduction, we might expect to find strong hierarchical relationships between males during the rut. With these hierarchies encoded in the sociality matrix then the dominant males would be labeled as strong structural leaders and the weaker males would be members of various reachability sets. In the same example, if a nursing offspring closely followed their mother, then the mother would exhibit structural leadership over her calf. Finally, note that while the mother is a structural leader to the calf, she may be influenced by a dominant male; making this mother a structural leader and a follower simultaneously and making the male an indirect structural leader of the calf. Therefore a binary classification of `leader vs.\ follower' is generally not appropriate.}

\begin{figure}[tb]
\includegraphics[width=64mm]{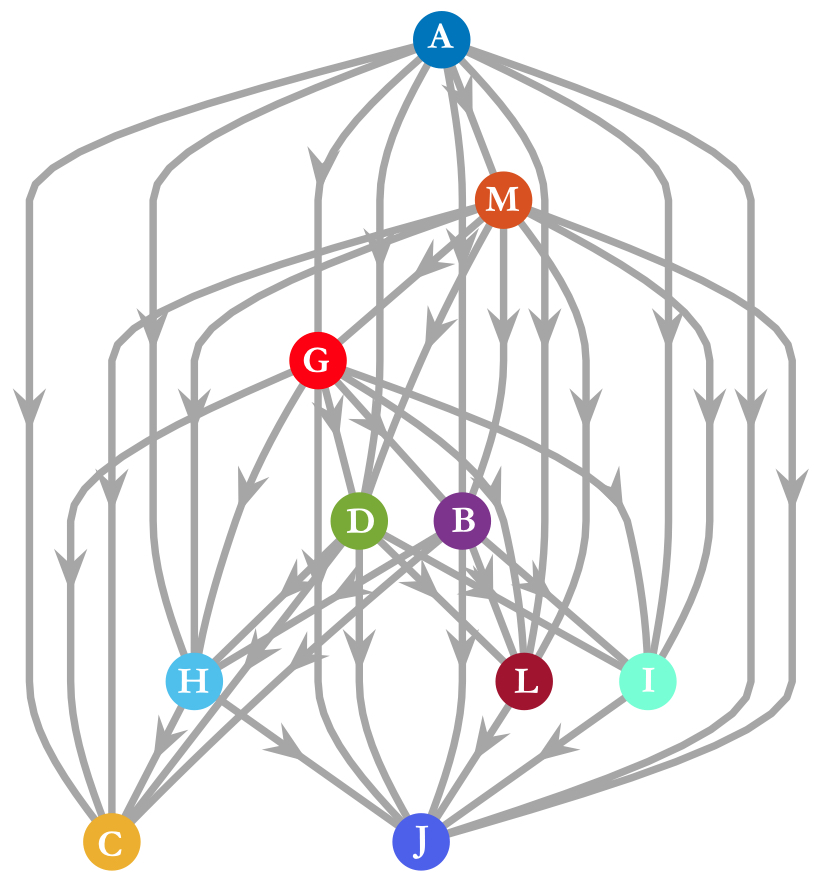}
\caption{Hierarchical Leadership in Pigeon Flocks. Here each directed edge represents the capacity for an individual to lead as defined by Nagy et al.\cite{nagy2010hierarchical}. For example, $L$ has the ability to lead $J$, and $J$ has the ability to lead no one.}\label{fig:hierarchy}
\end{figure}

\response{To further illustrate this point, consider the canonical example of hierarchical dynamics in pigeon flocks from Nagy et al.\cite{nagy2010hierarchical} depicted in Figure~\ref{fig:hierarchy}. In this example, nodes $C$ and $J$ have no structural leadership capacity as they have empty reachability sets. All other nodes however have the capacity to lead at least one other individual and thus all have some degree of structural leadership capacity. Notice, that with the exception of node $A$, each of the remaining individuals both lead and follow, i.e., they have non-empty reachability sets and are also members of others reachability set. The strength of their structural leadership would roughly mirror their vertical position in Figure~\ref{fig:hierarchy}.}

\response{Structural leadership is simply the capacity for a member of an animal society to lead other members of that society as dictated by the societies rules. In this sense structural leadership should really be seen more as a necessary but not sufficient condition for leadership to occur within a mobile animal group. However, in reality this component of leadership is quite important because it encodes the potentially important heterogeneity in interactions between specific pairs of individuals and more generally any hierarchies in the group.}

\medskip
\noindent\textbf{Informed Leadership.}
Informed leadership arises when a subset of the group are differentially informed and motivated to act on that information, e.g., a subset of the group senses a resource\cite{strandburg2013visual,reebs2000can}, or has information about a migration route\cite{mueller2013social,berdahl2018colnavrev}. Such leaders may be anonymous\cite{couzin2005effective}, or may indicate that they have information, for example by changing speed\citep{schultz2008mechanism} or signaling\cite{torney2011signalling}. 

In the case of our migrating caribou, both the experienced individuals leading the long-scale migration movement, and the individuals responding to local food and predation cues provide complementary examples of \textit{informed} leadership. 

\response{ Informed leadership generally arises from some underlying intent or motivation e.g., hunger or fear. For this reason, while the concept of informed leadership is intuitively sensible, from a mathematical standpoint it is both difficult to define and perhaps impossible to accurately infer without additional knowledge of the system.}

\medskip
\noindent\textbf{Target-Driven Leadership.}
Target-driven leadership is a specific subset of informed leadership. A target-driven leader is an informed leader (``informed by target") that uses a series of deliberate control inputs such as calls, explicit motions, etc.\ to guide a group toward a particular target state or set of target states. 
However, not all informed leadership is target-driven. 
For example, when an individual from a group of animals detects a predator, that individual becomes ``informed'' and tries to move away, and such abrupt change of motion may cause the rest of the group to follow. In this case, the first-reacting individual exhibits informed-leadership, but its sole ``target'', if any, is to move away from the predator instead of trying to lead the entire group away from the predator.

To be more precise, we characterize a target-driven leader as an individual that not only influences the group, but deliberately controls the group toward some target state. In addition, the removal of such an individual should result in the group not going towards the target state. Mathematically, we define this component  as follows. \response{Given that $\mathcal{A}$ is a set of target states, then individual $i$ is a target-driven leader (with respect to $\mathcal{A}$) if the net influence of $i$ on the group (see Eq.~\ref{eq:influence}) is nonzero and}
\begin{equation}\label{eq:targetstate}
\bm{y}(t)\rightarrow\mathcal{A}~\mbox{as}~ t\rightarrow\infty.
\end{equation}
That is, the individual directly influences the group as a whole and that influence results in the group progressing toward the target states.  

An example of a target-driven leader is a sheep dog. These dogs runs behind a group of sheep and through an intentional series of signals such as barking, eye contact and body posture the dog deliberately controls the sheep herd toward a given target state such as a barn or field.

\medskip
\noindent\textbf{Emergent Leadership.}
Asymmetrical influence, and thus leadership, may arise from social interactions rules alone, in the absence of social structure or differential information; we term this emergent leadership. This would be the case if animals used anisotropic social interaction rules. For example when individuals are more influenced by other individuals that are in front of them, then individuals in more frontal positions of the group are more influential, even if they have no additional information, motivation or status. Such \textit{emergent} leadership has recently been shown to be the case in our migratory caribou example\cite{torney2020inferring}.

Alternately, if individuals are more influenced by faster-moving group-mates\cite{pettit2015speed}, then those faster-moving individuals will have more influence. If those individuals are moving more quickly in response to information, or to signal dominance, then this would be informational or structural leadership, respectively, but if the increased speed is purely a function of the group dynamics, this would be an example of emergent leadership.

\subsection{Characteristics of Leadership}

\noindent\textbf{Distribution of Leadership.}
In animal groups decisions range from full distributed among all group members (`democratic') to dominated by a single or a few individuals (`despotic') \cite{conradt2003group,strandburg-Peshkin2018inferring}. 
It can be informative to quantify the number of individuals involved in a leadership role within the group. Similar to \cite{strandburg-Peshkin2018inferring} we refer to this as the \textit{distribution of leadership} which we define on a continuum that lies between \textit{centralized} and \textit{distributed} leadership.

At the scale of the entire herd, we might expect our migrating caribou to fall somewhere on this spectrum, bookended by primate societies with an alpha individual on one end and leaderless fission-fusion fish schools on the other. If we consider the mother-calf pairs as subgroups, we would expect the mother to be a \textit{centralized} leader. However, in a larger group containing many such pairs, we would expect \textit{distributed} leadership shared between the mothers. \response{The pigeon example in Figure~\ref{fig:hierarchy} illustrates that many systems fall somewhere between these two extremes. In this example nearly all of the individuals have some influence, yet it has a clear hierarchy so it is not fully decentralized; it therefore lies somewhere between centralized and distributed.}

\medskip
\noindent\textbf{Temporal Scale of Leadership.}
A leader may not be actively influencing the motion of other agents at all times and it is thus useful to quantify and understand the time scales for which a leader qualifies as a leader under any of the components of leadership.  Here, we consider two notions of time scales---consistency and granularity. For the following discussion consider dynamics of individuals, represented by discrete-time observations $\{x_i(t)\}_{t=0}^{T}$. 

\textit{Consistency} of leadership is simply defined as the proportion of the observation window for which a leader qualifies as a leader. More specifically, we classify leaders as {\it persistent} over the observation window if it is identified as a leader for the entire time window. Conversely, we classify a leader as {\it ephemeral} if it only qualifies as a leader for some small time window $[t,t+\tau]$, with $\tau\ll T$. A similar temporal leadership scale is presented in \cite{strandburg-Peshkin2018inferring} which ranges from variable to consistent but attempts to capture the same notion.

The \textit{granularity} of leadership concerns the resolution of time steps for which an individual acts as a leader. For example, a leader for daily activities might be different from one that is for seasonal activities. We can check for granularity by altering the time step we examine the dynamics under. In particular, quantify leadership using only the observations $\{x_i(kt)\}_{t=0}^{T/k}$ ($k>1$) for a large range of $k$. If a leader only acts on a coarse basis then they may not register as a leader for small $k$ but may then register as a leader for some larger $k$. In contrast a fine-scaled leader may register for many $k$.

In our migrating caribou example, the experienced individuals leading the broad migration path exhibit leadership that is \textit{persistent}, but perhaps has \textit{coarse granularity}. In contrast, the leadership of those animals responding to resources or predation threats along the way is \textit{ephemeral} and has \textit{fine granularity}.

Time scales present several challenges when attempting to infer leadership roles from a time series. If the granularity or observation window length do not match the natural time-scales of leadership then leadership events may be completely missed or misclassified. For example, consider a structural leader $\ell$ with the property that $I(x_i(t^-);y(t)|x_{\bar{i}}(t^-))=0$, i.e., a structural leader that does not directly influence the group--although it has the potential to. Regardless of the inference method such a potential leader will always be misclassified. Similarly consider an informational leader that only leads when they are within some radius to a known resource. Say that this event only occurred for a very short time window $[t,t+\tau]$, with $\tau\ll T$. If you only consider leaders that lead for the entire observation window, most aggregate measures will wash out such an ephemeral leadership event. For these reasons, carefully considering both consistency by studying sub samples of the data set as well as granularity by down sampling the data and retesting one will be able to obtain a much clearer picture of the leaders that are present in a mobile animal group.

\medskip
\noindent\textbf{Reach of Leadership.} 

\begin{figure}[tb]
\includegraphics[width=64mm]{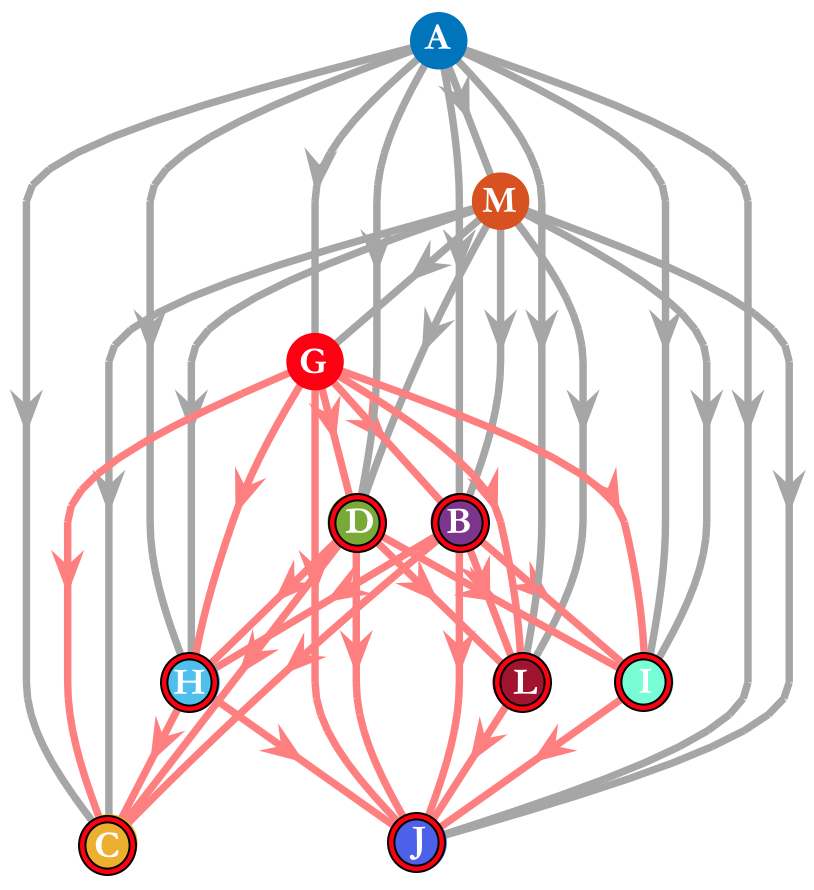}
\caption{Reach of Agent $G$. Each node with a red circle is within the reachability set and thus the reach of agent $G$.}\label{fig:reach}
\end{figure}
The \textit{reach} of \response{a leader quantifies the members of the group that the leader has potential influence over, directly and indirectly through subsequent interactions. Formally, we define the  reach of a leader as the members of that leaders reachability set on a graph associated with a particular source of leadership. In particular, let $\mathcal{G}$ be a graph where there exists a directed edge from node $j$ to node $i$ if $j$ has the capacity to lead $i$, where capacity to lead may be structural, emergent or informed leadership. Then the reach of agent $i$ is the reachability set of $i$ on $\mathcal{G}$.}

\response{Consider Figure~\ref{fig:reach}, where the graph represents the potential for structural leadership. In this example, individual $G$ has a reachability set of $\{D, B, H, L, I ,C, J\}$ and thus those 7 agents are within the \textit{reach} of structural leader $G$. Reach naturally lies on a continuum between \textit{local} and \textit{global}. If an agent exemplifies some form of leadership over all individuals this would be global reach; if an individual only leads some small subset of the group then this leader is considered local. In Figure~\ref{fig:reach} Agent $A$ has \textit{global reach} and agent $I$ has \textit{local reach}. }

In the case of our migrating caribou, the experienced migrants leading the entire herd on its broad migration path, would have \textit{global reach}, while the mother leading her calf on a finer-scale would have \textit{local reach}.

\medskip
\noindent\textbf{Observability of Leadership.} When we observe an animal society we do so imperfectly, mainly in two ways. First, any observed quantity is subject to noise and measurement errors. Secondly, and perhaps more importantly, there may be elements of the society which go {\it unobserved}. Such hidden variables and states may in turn act in our interpretation of leadership. In fact, the strongest leadership might not be detectable if the data are not appropriate. Across various taxa, leaders may use vocal cues\cite{stewart1994gorillas,fossey1972vocalizations}, gestures\cite{smith2003animal} or movements that are too fine to be picked up by GPS (e.g.  pre-flight flapping\cite{black1988preflight}) to initiate or control movement. If the resulting movement is synchronized, leadership inference based on trajectories will fail. Worse, if in the resulting movement, the least dominate individuals respond first to the cues, it could appear as though those individuals are leading.

In the case of our migrating caribou, lead animals may stand up to signal departure, or motivate others to start moving. This would not be captured by GPS tags and so would be \textit{hidden} to inference methods based on trajectories alone.

Quantification of hidden leadership in practice is quite difficult by definition. Namely, if you have detected leadership it was observed. Doing this in theory however is quite trivial. As defined in Section~\ref{sec:math-frame} we define the full system dynamics via $x$, $y$, $S$ (and/or some mix of these). When the system is being observed, the observed variables, denoted $\hat{\bm{x}}$, $\hat{\bm{y}}$, and $\hat{S}$, can differ from the true ones.
\response{We term an individual's leadership role `hidden' if it exhibits leadership defined in terms of the intrinsic variables $(\bm{x},\bm{y},S)$ but does not appear to do so given the observed variables $(\hat{\bm{x}},\hat{\bm{y}},\hat{S})$; a leader that is not hidden is then called an observable leader.}

\subsection{Real World Animal Behavior and the Anatomy of Leadership}\label{sec:sandbox2}

Here we discuss real world animal interactions, and we do so in a manner to emphasize the terminology of our anatomy of leadership taxonomy.

We expect to find structural leadership in relatively stable animal groups, often having complex social hierarchies, such as cetaceans, wolves, wild dogs, elephants and primates \cite{king2008dominance,brent2015ecological,payne2003sources,peterson2002leadership,lusseau2009emergence}. The canonical example is the so-called `alpha' individual in a primate society, who has some level of control over an entire group over a long period of time (assuming he society is stable)\cite{flack2006policing}. In our taxonomy, this dominant individual would be a \textit{persistent}, \textit{centralized}, \textit{structural} leader with a \textit{large reach}. It is important to note that in such societies, structural leadership may be well correlated with informational leadership. For example, a matriarch elephant may have better information about rarely visited water holes, as well as greater per-capita influence to lead her group to them.

We expect informational leadership to dominate in animal groups composed of unrelated individuals and unstable membership (i.e., fission-fusion dynamics), such as fish schools and bird flocks. A single arbitrary member of a fish school may perceive a respond to a threat, causing those around it to also startle, or the entire group to make an evasive maneuver \cite{rosenthal2015revealing}. This is an example of \textit{centralized}, \textit{ephemeral}, \textit{informational} leadership with a \textit{limited or global reach}, depending how much of the group responded. Similarly, some fraction of the same school might have information about where or when a food resource might occur and lead the entire school to that time-space location\cite{laland1997shoaling,reebs2000can,strandburg2013visual}. In our terms those fish are \textit{distributed}, \textit{ephemeral}, \textit{informational} leaders with \textit{global reach}.

Informational leadership is also common for movement at long length scales. In flocks of pigeons, better informed individuals act as leaders during homing flights \cite{flack2012leaders}. (However, it should be noted that pigeons also exhibit a structural hierarchy too\cite{nagy2010hierarchical}.) During migratory movements, older, more experienced, birds guide groups on efficient migration routes \cite{mueller2013social,berdahl2018colnavrev}.  In both of these examples the informed birds are \textit{centralized}, \textit{persistent}, \textit{target-driven}, \textit{informational} leaders with \textit{global reach}.

In migrating white storks some individuals actively seek thermals updrafts, which are necessary for them to get efficient lift to complete the migration, while others tend to copy, by moving towards individuals who are already in thermals \cite{nagy2018}. This is a specific example of a general phenomenon, emergent sensing\cite{berdahl2018colnavrev}, in which a group spans an environmental gradient and individuals in the `preferred' end of the gradient alter their behaviour (purposefully or not) in a way that cause the entire group to climb the gradient \cite{berdahl2013emergent,torney2011signalling}. In general such leadership would be \textit{distributed} and \textit{ephemeral} (although could be \textit{persistent} if, like in the storks, the same individuals always find the thermals) \textit{informational} leadership with \textit{global reach}.

\section{A Model Sandbox for Validating Leadership-Inference Methods}\label{sec:sandbox}
\response{Ultimately one would like to develop methods to infer and classify leadership from empirical data. This is of course a long-standing and non-trivial challenge, and a pragmatic approach is to first test inference methods on simulated data where the leadership type and distribution is known because it is programmed in explicitly. For mobile animal groups an obvious starting point is to modify classic flocking/schooling/herding models (e.g., \cite{reynolds1987flocks,vicsek1995novel,couzin2002collective}) to include known leadership structures. In this section we first describe a canonical model of collective motion -- the so-called zonal model\cite{couzin2002collective,couzin2005effective}. Following that, we modify the model to incorporate the various leadership sources and characteristics described in this paper.
}

\subsection{Basic Collective Motion Model}

\begin{figure}[htb]
\includegraphics[width=64mm]{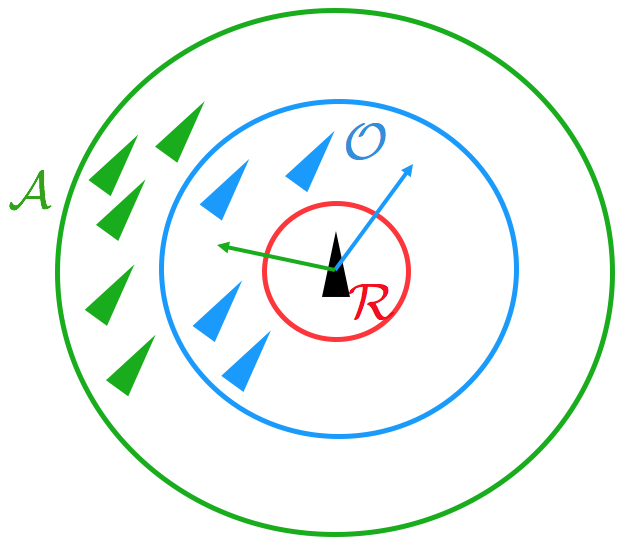}
\caption{Schematic of Zonal Flocking Model. The black triangle is the focal individual. The red ring demarks the zone of repulsion $\mathcal{R}$. The blue circle is the orientation zone $\mathcal{O}$, the focal individual attempts to align with the agents in this zone (blue triangles in the figure). The outer ring is the attraction zone $\mathcal{A}$ and the focal individual attempts to get closer to these agents (the green triangles in the picture). The resulting desired direction is then the sum of the green and blue vectors. }\label{fig:zon-model-illustration}
\end{figure}

\response{Following Couzin et al.\cite{couzin2002collective,couzin2005effective},} for each agent, numbered $i=1,..,N$, and each time $t$, a position vector $\bm{c}_i(t)$, a direction vector $\bm{v}_i(t)$ and a speed $s_i$ are maintained. At each time step agent $i$ computes a \textit{desired direction} $\bm{d}_i(t)$ based on neighbors in three different zones, depicted in Figure~\ref{fig:zon-model-illustration}. 

The first zone to consider is called the \textit{repulsion} zone and is denoted by $\mathcal{R}$. This zone ensures that `personal space' is maintained for each agent. If any other agent is in the repulsion zone $\mathcal{R}$, for focal individual $j$, then the desired direction in the next time step is defined by
\begin{equation}\label{eq:repel}
\bm{d}_i(t+\Delta t) = - \sum_{
\substack{j\not=i \\ j \in \mathcal{R}}}\frac{\bm{c}_j(t)-\bm{c}_i(t)}{|\bm{c}_j(t)-\bm{c}_i(t)|}.
\end{equation}
This desired direction ensures that a collision will not occur at time $t+\Delta t$. However, if for the focal individual $\mathcal{R} =\emptyset$ then the focal individual attempts to get closer to agents in their \textit{attraction} zone $\mathcal{A}$ and orient with agents in their \textit{orientation} zone $\mathcal{O}$. This is accomplished by choosing a desired direction at time $t +\Delta t$ in the following way:
\begin{equation}
\begin{aligned}
\bm{d}_i(t+\Delta t) &= \alpha\sum_{\substack{j\not=i \\ j \in \mathcal{A}}}\frac{\bm{c}_j(t)-\bm{c}_i(t)}{|\bm{c}_j(t)-\bm{c}_i(t)|}\\
&+(1-\alpha)\sum_{\substack{j\not=i \\ j \in \mathcal{O}}} \frac{\bm{v}_j(t)}{|\bm{v}_j(t)|}.
\end{aligned}
\end{equation}
Where $\alpha$ is a parameter that controls the relative strength of attraction and alignment. \response{For example, a flock of geese -- dominated by alignment -- would have a relatively low $\alpha$, while a swarm of insects -- dominated by attraction -- would have a relatively high $\alpha$.}

\response{The desired direction vector, $\bm{d}$, is normalized to a unit vector $\bm{\hat{d}}(t+\Delta t) = \frac{\bm{d}_i(t+\Delta t)}{|\bm{d}_i(t+\Delta t)|}$. Next, to represent uncertainty stemming from limitations of sensory and cognitive abilities, this unit vector is transformed into $\bm{d}''(t+\Delta t)$ by rotating it by a small angle drawn from a circular-wrapped Gaussian distribution centered at zero. Finally, it is assumed that individuals can turn at a maximum rate of $\theta$ radians per unit time.  Therefore, if the difference between an individual's current direction, $\bm{v}_i(t)$, and its desired direction for the next time step, $\bm{d}''_i(t+\Delta t)$, is less than $\theta \Delta t$ then the desired direction is achieved and $\bm{v}_i(t+\Delta t) = \bm{d}''_i(t+\Delta t)$. Otherwise, that individual's direction $\bm{v}_i(t+\Delta t)$ is the result of rotating $\bm{v}_i(t)$ by $\theta \Delta t$ radians towards their desired direction $\bm{d}''_i(t+\Delta t)$.}

After the heading is assigned, the position at $t+\Delta t$ can be computed by 
\begin{equation}\label{eq:basicflocking}
\bm{c}_i(t+\Delta t) = \bm{c}_i(t)
+\bm{v}(t+\Delta t)s_i\Delta t,
\end{equation}
where $s_i$ is the speed of individual $i$.

\subsection{Explicitly Adding Sources of Leadership}
\response{While this base model captures a wide variety of flocking, swarming and schooling behavior it does not account for leadership explicitly. In order to explicitly test leadership inference methods, it is helpful to make a few  simple modifications to this base model: (1) add a sociality matrix\cite{bode2011impact} (structural leadership) (2) add ``informed" individuals to the group\cite{couzin2005effective} (informed leadership) and (3) make interaction rules isotropic\cite{couzin2002collective} (emergent leadership).}
\\

\noindent\textbf{Structural Leadership}\response{ To add structural leadership we introduce} a sociality matrix $S=[S_{ij}]_{N\times N}$.
$S_{ij}\not=0$ if agent $i$ can be influenced by agent $j$.
\response{More generally, $S_{ij}$ is a continuous value that gives the relative influence of individual $j$ on individual $i$.}
To take this into account the desired-direction computation is modified to \response{weight the influence of each neighbor relative to $S_{ij}$} (rather than \response{an equal weighting of everyone} in $\mathcal{A}$ and $\mathcal{O}$, i.e.,
\begin{equation}
\begin{aligned}
\label{eq:flockingwalpha}
\bm{d}_i(t+\Delta t) &=& \alpha\sum_{\substack{j\not=i \\ j \in \mathcal{A}}}S_{ij}\frac{\bm{c}_j(t)-\bm{c}_i(t)}{|\bm{c}_j(t)-\bm{c}_i(t)|}  \\
&+&(1-\alpha)\sum_{\substack{j\not=i \\ j \in \mathcal{O}}} S_{ij} \frac{\bm{v}_j(t)}{|\bm{v}_j(t)|}.
\end{aligned}
\end{equation}
\response{Adding this sociality matrix to the base model allows for structural leadership to be explicitly built in. This is an advantage as you can then see if \textit{post-facto} if the structural leadership placed in the model can be extracted by a candidate inference method.}
\\

\noindent\textbf{Informed Leadership}\response{
To simulate informed leadership, a subset of the agents are} given knowledge of a preferred direction $\bm{g}$ (more generally each informed agent is given their own not necessarily equal preferred direction $\bm{g}_i$)\cite{couzin2005effective}. This preferred direction may be part of a migration route or the direction of a prey or known resource. Non-informed group members, have no knowledge of $g$ and may or may not know which individuals are informed. Following Couzin et al.\cite{couzin2005effective}, to integrate this into the model the informed individuals balance between the social interactions and their preferred direction with a weighting term $\omega$. In particular, informed individuals have a desired dircetion $\bm{d}'$, given by
\begin{equation} \label{eq:zonalwinfo}
\bm{d}'_i(t+\Delta t) = \frac{\bm{\hat{d}}_i(t+\Delta t) + \omega \bm{g}_i}
{|\bm{\hat{d}}_i(t+\Delta t) + \omega \bm{g}_i|}.
\end{equation}

If $\omega=0$ the preferred direction is completely ignored and only social interactions are followed. As $\omega$ increases toward 1 the influence of the preferred direction is balanced with influence of the social interactions. With $\omega > 1$ the preferred direction is favored over social interactions.
\\

\noindent\textbf{Emergent Leadership}
\response{One way to make a test case for inferring emergent leadership, is to make interactions spatially asymmetric. In particular, one can simply add `blind zones'\cite{couzin2002collective} to the model described in Eqs.~\ref{eq:repel}-\ref{eq:basicflocking}. In this case the zones $\mathcal{A}$ and $\mathcal{O}$ are missing wedges behind them and individuals in those wedges are ignored. If these blind zones are large enough, individuals are more influenced by individuals in front of them\cite{torney2020inferring}.}

\subsection{Testing Characterizations of Leadership}

\noindent\textbf{Distribution of Leadership}
\\
\response{Using the framework presented in the previous sections one can explore a variety of distributions of leadership, ranging from centralize to distributed\cite{strandburg-Peshkin2018inferring}. For structural leadership, the spectrum could range from a sociality matrix with a hub structure (centralized) to a one with a random dense connectivity, or even fully connected (decentralized). For informed leadership, 
the fraction of the group having a non-zero value of $\omega$ would roughly span the spectrum of the distribution of leaderhip.
We note that the distributions of structrural and informed leadership are potentially orthoginal. For example a group could have highly centralized structural leadership in tandem completely distributed informational leadership, or vise versa.
}
\\

\noindent\textbf{Temporal Scale of Leadership} \response{Temporal consistency and granularity of leadership can be built into this leadership model by making the model parameters associated with leadership time dependent, e.g, $[S_{ij}(t)]$, $\omega(t)$ and $\bm{g}(t)$. For example, one could remove or change the preferred direction at regular time intervals by defining time-varying $\omega(t)$ and then see if an inference algorithm could detect this change. }
\\

\noindent\textbf{Reach of Leadership} 
\response{By setting specific examples of the sociality matrix one can experiment with a variety of leadership reach scenarios and test the ability of various inference measures to recover them.}
\\

\noindent\textbf{Observability of Leadership}
\response{There is a vast set of variations that could be made to the framework presented here to encode potential for leadership to to be driven by non-trajectory based cues or signals\cite{brown1991food,stewart1994gorillas,fossey1972vocalizations,smith2003animal,black1988preflight}.  One obvious example (which is also ubiquitous in nature) is auditory signalling, which could provide long-range interactions\cite{torney2011signalling}.
}
\\

\subsection{A Potential Pitfall: Influence vs.\ leadership}
\response{Consider a mobile-animal group where each member is governed by Eq.~\ref{eq:basicflocking} and the direction is decided by Eqs.~\ref{eq:repel} \& \ref{eq:flockingwalpha}. Furthermore, define $S_{i(i+1)}=1$ and 0 otherwise for $i\in \{1,\dots, N-1\}$ and let $\alpha\in[0,1]$. This describes a simple chain topology, where each individual has the capacity for structural leadership over at most one other agent. In particular, each agent orients and attracts to (follows) at most one other agent in the group. However, it is important to note that every agent avoids collisions with all other agents (the sociality metrix applies to Eq.\ \ref{eq:flockingwalpha}, but \textit{not} Eq.~\ref{eq:repel}). } 

\response{In this example the incidental social interactions, such as those caused by repulsion, cause a real problem for the majority of influence/causal inference algorithms. For example, if one blindly applied optimal causation entropy\cite{sun2015causal} or transfer entropy\cite{Schreiber2000} to infer who leads whom then these algorithms would conclude an all-to-all leadership graph. By construction however we know this is incorrect and that the underlying influence graph is a simple chain. The issue here is that these measures\cite{sun2015causal,Schreiber2000}, and causal inference from information in general, are not explicitly measuring leadership but reductions in uncertainty about a particular variable. In this example, the minor local repulsion interactions cause enough ``information flow" over time to trigger these algorithms/measures. However, as discussed in Appendix~\ref{sec:background} conflating influence, information flow, causality and leadership is a non-trivial challenge, which is nicely highlighted by the present example.}

\section{Afterward}

Traditional approaches to leadership inference have focused on a single defining characteristic, e.g., position within a group, social hierarchy, information flow or influence. We believe that, in general, none of these concepts alone fully captures leadership. In this manuscript we have begun to show that a multifaceted approach where multiple axes of leadership are analyzed provides a more complete classification of the leadership structure. This formalism should serve to link questions about empirical systems with the appropriate analytical tools to address those questions.  While this taxonomy we provided is surely not complete we hope that this effort will serve as a starting point for formalizing a multifaceted approach to leadership inference. 

\response{Multiple technological advances in sensors, computer vision have led to the availability of more high-resolution collective motion data than ever before\cite{hughey2018challenges}. As such the near future is an opportune time to make meaningful advances in leadership inference. Causal inference and information theory show a lot of promise in this arena but as we have shown throughout this manuscript leadership is a highly intricate and multifaceted subject and neither causal inference nor information theory may be up for the task alone. We hope that as new inference algorithms come to be the formal language and toy models developed here will serve as a proving ground. We believe that being able to carefully classify the components of leadership being inferred will be invaluable for practitioners and theorists as they begin to tackle all the high-resolution data as it becomes available.}

\section{Acknowledgments}

JG and AMB were supported by Omidyar Fellowships from the Santa Fe Institute.  
AMB was also supported a grant from the John Templeton Foundation.  EB and JS were in part by the Army Research Office grant W911NF-16-1-0081, and JS by the Simons Foundation grant 318812, and EB by the Office of Naval Research grant N00014-15-1-2093. The opinions expressed in this publication are those of the authors and do not necessarily reflect the views of the John Templeton Foundation.

\bibliography{leadership}

\newpage
\appendix

\section{Information Flow, Causality, Influence and Leadership}\label{sec:background}
 Information theory provides sophisticated measures for rigorously quantifying concepts like ``the reduction in uncertainty about the present state of $X$ given past states of $Y$." As such, these measures are often associated with concepts like information flow, causality, influence and even leadership---and often all of these terms are used interchangeably.  These measures are often viewed as less subjective inference methods because almost no assumptions need to be made about the structure of the system being observed. As a result, information theory has become a popular tool for inferring leadership from time series\cite{lord2016inference,butail2016model,sun2015causal, nagy2010hierarchical,akos2014leadership,jiang2017identifying,watts2017validating,ye2015distinguishing,darmon2015followers}.
However, while influence, information flow and causality are all closely related to the notion of leadership, these concepts are inherently different and therefore are not readily interchangeable. Furthermore, recent work has begun to show that these information measures fail to even capture information flow\cite{james2016information} let alone leadership.

The following appendix discuses information flow, causality and influence and provides motivation for why we do not believe any of these alone fully quantifies leadership.

{\bf Information flow and entropy,} as we have argued in previous mathematical works \cite{bollt2013applied, bollt2012synchronization,lord2016inference}, is a fundamental concept in coupled (dynamical) systems, and the associated  stochastic processes.  Information theory, as formulated upon Shannon entropy and its variants, basically describes the average ``surprise'' one should attribute to observing a specific value or state of a random variable. More formally, such quantification of surprise or (un)predictability is referred to as ``entropy'' and can be defined rigorously as a function of the underlying probability distributions. When the time evolution of multiple variables are considered, the state of a variable often depends on the history of a set of related variables, and such inter-variable dependencies can be viewed as ``information flow''. Explicit characterization of information flow in coupled systems can be done by quantifying how informative (again as a notion of surprise) one should be in measured observations conditioned on given previous observations, giving rise to commonly used measures such as transfer entropy~\cite{Schreiber2000} and causation entropy~\cite{sun2014causation, sun2015causal, cafaro2015causation}. In other words, information flow describes the reduction in uncertainty regarding forecasts for predictions associated with conditioning on the past in various combinations.  Thus whether by Granger causality\cite{granger1980testing},  transfer entropy\cite{Schreiber2000}, causation entropy\cite{sun2014causation, sun2015causal, cafaro2015causation,sun2014identifying}, or some other method, the idea is to ask if there is a reduction in uncertainty with knowledge of the past of a perhaps coupled variable.  Clearly, this question is universally relevant from a wide range of scientific fields of science or mathematics.    However, part of the theme of this paper is that these information flow concepts themselves are not sufficient or equivalent as leadership.
 
{\bf Causation} is a related but not identical concept as information flow. The notion of causality has many interpretations,  depending on the context, from philosophical \cite{russell1913mysticism,russell1948human, bigelow1990metaphysics}, to statistical \cite{pearl1995theory, pearl2011bayesian, pearl2009structural, skyrms2012causation, freedman2005linear}, to dynamical \cite{granger1980testing, barnett2009granger,Schreiber2000, sun2014causation}.  Here we will avoid the philosophical direction  entirely, but note that some of these do coincide with the others.  Statistical perspectives are sometimes relevant to a stochastic process, especially from the influential work of Pearl \cite{pearl1995theory, pearl2011bayesian, pearl2009structural}, associated with a calculus for understanding interventions, but not always relevant to our context.  We are more so interested in understanding interpretations of causal influence, of a free running system, that is, a system that is passively observed rather than actively probed. As such, this relates more closely, almost synonymously to the concepts of information flow in a stochastic process, but not quite identically.  We take the same perspective as Granger in his line of reasoning that eventually lead to the 2003 award of the Nobel Prize in Economics; Granger's fundamental principles were that 1) cause happens {\it before} effect, and 2) a cause necessarily contains unique information concerning future states of its effect~\cite{granger1980testing}. In details the so-called Granger causality is a specific computation that assumes a linear stochastic process, and as such, it was shown \cite{barnett2009granger} to be entirely equivalent to transfer entropy computed by other means (in information theoretic by the Kullback-Liebler divergence appropriately conditioned) in the special case of a linear stochastic process with Gaussian noise.  So said, while the underlying principles of Granger are the same, the details of computation may differ.  

{\bf Influence} can now be described within this formalized framework as related to, but somewhat distinct from  leadership, depending on if we are relating interactions between agents in terms of information theory, reduction of uncertainty, or some other underlying principle, including the potential goal of controlling the system.  Consider that some agents in a group may be leaders, with various ways to interpret this phrase to be stated subsequently below.  A measure of leadership may be associated with information flow for example, or as a proxy for causal influences that leaders may change states, before other agents, a concept which will follow analogously to cause that comes before effect.  An influential member of a group is not necessarily a leader, although in some sense influence is a kind of leadership \textit{de facto} in the sense that influence is comparable to the possibility to cause others to change their behavior (dynamics).  

So said then what is the difference between influence, causation, and leadership, from the perspective of information flow?  In some interpretations then, influence or causation over others and leadership are almost synonymous but with important distinctions.   When leadership is viewed through the lens of reduction of uncertainty (thus measurable by causation inference and information flow), then causation and influence becomes a synonym for leadership. Therefore, if a leadership action is active and observable, then causation and information flow are relevant concepts that enable one to define and empirically score the leadership.  However, there are other notions of leadership that are clearly beyond the scope of information flow.  Herein, by using a taxonomy of leadership, we expand beyond the typical causation and information flow concepts~\cite{lord2016inference,butail2016model, strandburg2013visual} to allow for those features which may be missed through the narrow interpretation of entropy, including   structure, degree to which agents are informed, distribution, time and space scales, and target-drive are some of the other aspects that we will discuss here.

\end{document}